\documentclass{ifacconf}

\usepackage{graphicx}      
\makeatletter
\let\old@ssect\@ssect 
\makeatother
\usepackage{natbib}        

\usepackage{latexsym}
\usepackage{amsfonts,amssymb,amsmath}
\usepackage{hyperref}
\makeatletter
\def\@ssect#1#2#3#4#5#6{%
  \NR@gettitle{#6}
  \old@ssect{#1}{#2}{#3}{#4}{#5}{#6}
}
\makeatother

\usepackage{xcolor}
\usepackage{algorithm}

\usepackage{algpseudocode}
\def\BibTeX{{\rm B\kern-.05em{\sc i\kern-.025em b}\kern-.08em T\kern-.1667em\lower.7ex\hbox{E}\kern-.125emX}}
\usepackage{subcaption}
\usepackage{comment}

\usepackage{dcolumn,booktabs,adjustbox}
\newcolumntype{d}[1]{D{.}{.}{#1}}

\newcolumntype{R}[2]{%
    >{\adjustbox{angle=#1,lap=\width-(#2)}\bgroup}%
    l%
    <{\egroup}%
}
\newcommand*\rot[2]{\multicolumn{1}{R{#1}{#2}}}

\begin{document}
\begin{frontmatter}

\title{Data-driven Steering Torque Behaviour Modelling with Hidden Markov Models} 


\author[First]{Robert van Wijk}, 
\author[Second]{Andrea Michelle Rios Lazcano}, 
\author[Second]{Xabier Carrera Akutain}, 
\author[First]{Barys Shyrokau} 

\address[First]{Delft University of Technology, 
   Delft, Netherlands (e-mail: \{B.Shyrokau\}@tudelft.nl).}
\address[Second]{Toyota Motor Europe, 
   Zaventem, Belgium (e-mail: \{Andrea.Lazcano, Xabier.Carrera.Akutain\}@toyota-europe.com)}
   
\begin{abstract}                
Modern Advanced Driver Assistance Systems (ADAS) are limited in their ability to consider the driver's intention, resulting in unnatural guidance and low customer acceptance. In this research, we focus on a novel data-driven approach to predict driver steering torque. In particular, driver behavior is modeled by learning the parameters of a Hidden Markov Model (HMM) and estimation is performed with Gaussian Mixture Regression (GMR). An extensive parameter selection framework enables us to objectively select the model hyper-parameters and prevents overfitting. The final model behavior is optimized with a cost function balancing between accuracy and smoothness. Naturalistic driving data covering seven participants is obtained using a static driving simulator at Toyota Motor Europe for the training, evaluation, and testing of the proposed model. The results demonstrate that our approach achieved a 92\% steering torque accuracy with a 37\% increase in signal smoothness and 90\% fewer data compared to a baseline. In addition, our model captures the complex and nonlinear human behavior and inter-driver variability from novice to expert drivers, showing an interesting potential to become a steering performance predictor in future user-oriented ADAS.
\end{abstract}


\end{frontmatter}


\section{Introduction}

Driver Steering Assistance Systems (DSAS) can (partially) take over the vehicle's lateral control, thus effectively sharing control with the driver. However, commercial DSAS focus on path-tracking performance without considering the interaction with the driver. Integrating knowledge on the driver allows DSAS to better match driver intentions. However, modeling driver steering behavior is still a challenge due to the highly complex, stochastic, and variable human nature \citep{kolekar2018}.

State-of-the-art driver models can be categorized into parametric, non-parametric, and mixed approaches. Parametric approaches \citep{saleh2011,niu2020} provide an intuitive approximation of the steering behavior and are based on physical principles using a priori assumed accurate analytical models. Non-parametric approaches \citep{jugade2019} can capture the nonlinear behavior by inferring model structure from data without predefined assumptions. Mixed approaches \citep{lefevre2014} aim to combine driver's intuition with the ability to learn nonlinear behavior thanks to a model structure that partially follows a theoretical background with a data-driven design. However, the mentioned approaches (except parametric) are based on \textit{steering angle} behavior. Steering torque, which is required for driver-vehicle interaction, has not yet been investigated.  

The proposed work is inspired by \cite{lefevre2014} and focuses on the prediction of a \textit{continuous steering torque}, making the model appropriate for the development of a new haptic DSAS \citep{lazcano2021}. This paper is structured as follows. The proposed driver model and inference process are explained in Section \ref{sec:drivermodel}. Section \ref{sec:parameter-selection-framework} covers an extensive parameter selection framework to design model behavior, followed by a simulator experiment to validate the method on naturalistic driving data in Section \ref{sec:driving-simulator-experiment}. The results are presented in Section \ref{sec:results-discussion} and the conclusions are highlighted in Section \ref{sec:conclusion}.

\section{Driver Model}
\label{sec:drivermodel}
\cite{pentland1999} describe human steering behavior as a set of discrete (hidden) states, each with its unique control behavior. Sequencing the states together with a Markov chain, one obtains a Hidden Markov Model. \cite{lefevre2014} propose to use Gaussian normal distributions to learn the relation between scenario and steering angle. The proposed study suggests learning the relation with the steering torque instead of the steering angle. 

\subsection{Model Structure}
\label{subsec:hmm}
Driver steering behavior is modeled with a fully connected HMM. The aim is to learn the joint probability distribution between the driving scenario, defined as a vector of features, $F \in \mathbb{R}^{N_f}$, and the steering torque, $T_d$. 
The model is described by four parameters. First, the number of hidden states $K$ determines the model configuration. Second, the vector $\boldsymbol{\pi} \in \mathbb{R}^{K}$ determines the prior probability of starting in state $k$ at time $t=1$. Third, state transition matrix $\boldsymbol{A} \in \mathbb{R}^{KxK}$ determines switching behaviour, where each element $a_{jk}$ describes the probability of switching from state $j$ at time $t-1$ to state $k$ at time $t$. Finally, state control behaviour is defined by the set $B$, where each element $b_k$ describes the probability distribution $P(x_t|k)$ of being in state $k$ and observing the joint driving scenario and driver steering torque $x_t = [F_t,T^d_t]^{\top}$ at time $t$. As the driver steering torque is continuous, each distribution $P(x_t|k)$ is assumed as a single multivariate Gaussian, $b_k \sim \mathcal{N}(\mu_k,\Sigma_k)$, defined as,
\begin{equation}
    \label{eq:gaussiansplit}
    \mu_k = \begin{bmatrix} \mu_k^{F} \\ \mu_k^T \end{bmatrix} \text{ and } \Sigma_k = \begin{bmatrix} \Sigma_k^{FF} & \Sigma_k^{FT} \\ \Sigma_k^{TF} & \Sigma_k^{TT} \end{bmatrix}
\end{equation}

Denoting $s_t$ as the state at time $t$, let $\boldsymbol{S} = \{s_1,...,s_T\}$ represent the hidden state sequence from time $t=1$ to $t=T$. Together with $\boldsymbol{X} = \{x_1,...,x_T\}$, representing the observation sequence, the joint state-observation probability distribution $P\left(\boldsymbol{X},\boldsymbol{S}|\boldsymbol{\theta}\right)$ is defined in \cite{bishop2006} as,
\begin{multline}
    \label{eq:jpdf}
    p(\boldsymbol{X},\boldsymbol{S}|\boldsymbol{\theta}) = \\ p(s_1|\boldsymbol{\pi})\left[\prod^T_{t=2}P(s_t|s_{t-1},\boldsymbol{A})\right]\prod^T_{t=1}p(x_t|s_t,B)
\end{multline}

The model parameters $\boldsymbol{\theta} = \{\boldsymbol{\pi},\boldsymbol{A},B\}$ are learned from recorded driving data with the Baum-Welch Algorithm \citep{rabiner1989}. The feature set $F$ and the number of hidden states $K$ are determined in Section \ref{sec:parameter-selection-framework}.

\subsection{Model Inference}
\label{subsec:gmr}
At each timestep, the state control distributions $B$ represent a mixture of $K$ multivariate Gaussians. Therefore, estimation of steering torque $T_d^{est}$ is performed with Gaussian Mixture Regression (GMR) \citep{tian2011} as,
\begin{equation}
    \label{eq:gmr}
    T_{d,t}^{est} = \sum_{k=1}^{K}\alpha_{t,k}^{F}\left[\mu_k^{FF} + \Sigma_k^{TF}(\Sigma_k^{FF})^{-1}(F_t - \mu_k^{FF})\right]
\end{equation}

where $\alpha_{t,k}$ represents the mixture weights. \cite{calinon2010} proposed to calculate these weights recursively with the Forward Variable \citep{rabiner1989}, defined as,
%
\begin{equation}
    \label{eq:forwardvariable}
    \alpha_{t,k}^F = \frac{
            \left(\sum_{j=1}^{K}\alpha_{t-1,j}^{F}a_{jk}\right) \mathcal{N}(F_t | \mu_k^{F},\Sigma_k^{FF})
        }
        {
            \sum_{i=1}^{K}\left[\left(\sum_{j=1}^{K}\alpha_{t-1,j}a_{ji}\right)  \mathcal{N}(F_t | \mu_k^{F},\Sigma_k^{FF}) \right]
        }
\end{equation}

and corresponds to the probability of observing the partial observation sequence $\{F_1,F_2,...,F_t\}$ and being in state $k$, given the model parameters $\boldsymbol{\theta}$. The Forward Variable at time $t=1$ is initialized with prior probability $\pi_k$ as,
\begin{equation}
    \label{eq:initalpha}
    \alpha_{1,k}^{F} = \frac{\pi_k \mathcal{N}(F_1 | \mu_k^{F},\Sigma_k^{FF})}{\sum_{j=1}^{K}\left[\pi_j \mathcal{N}(F_1 | \mu_k^{F},\Sigma_k^{FF}) \right]}
\end{equation}

\section{Parameter Selection Framework}
\label{sec:parameter-selection-framework}
\begin{figure*}
    \centering
    \includegraphics[width=0.7\textwidth]{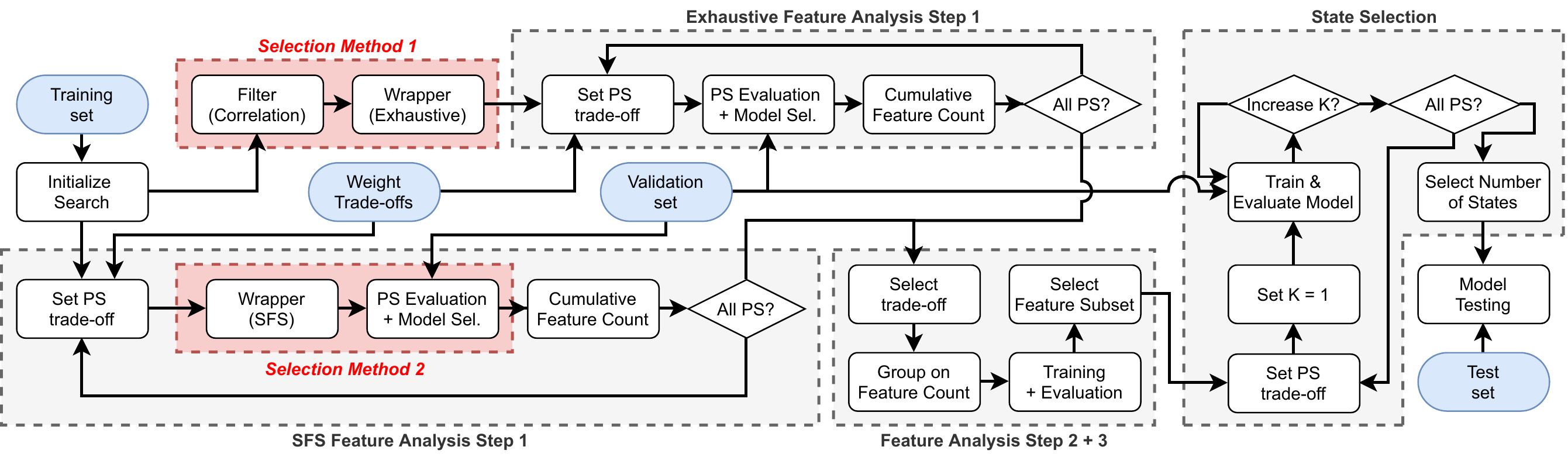}
    \caption{The proposed parameter selection framework as explained in \ref{subsec:feature-selection-methodology}, \ref{subsec:evaluation-and-selection} and \ref{subsec:model-complexity}}
    \label{fig:framework}
\end{figure*}

This section proposes a parameter selection and overfit prevention framework, Fig. \ref{fig:framework}, to objectively find the optimal feature set $F$ and the number of states $K$. The aim is to maximize estimation accuracy and generalization capabilities while minimizing oscillations. 

\subsection{Feature Selection Background}
\label{subsec:feature-selection-background}
Feature selection removes redundant features to improve learning speed, model interpretability, and model performance, while reducing the risk of overfitting and required data storage \citep{adams2019}.

The current work proposes an adaptation of the work by \cite{faller2016} for an HMM-based prediction of unintentional lane changes. Even though the HMM was used as a classifier, a similar supervised approach is adopted due to the promising results and lack of alternative regression-based strategies. The method consists of two wrapper-based selection methods, an exhaustive search and a sequential forward search (SFS). Wrapper methods determine inter-feature dependencies and redundancies by utilizing the training algorithm as a performance measure \citep{jovic2015}. This makes wrappers computationally expensive but allows them to obtain better performing subsets. Faller uses the SFS wrapper to validate the candidate subsets generated by the exhaustive wrapper. The current work adopts and extends the method by replacing the manual pre-selection with a filter method (\ref{subsec:feature-selection-methodology}). Filter methods are independent of the learning algorithm, computationally efficient, and rank features based on data performance metrics \citep{jovic2015}. Feeding the filter output to the exhaustive wrapper ensures that redundant features are removed. Additionally, a weighted performance score (\ref{subsec:evaluation-and-selection}) is implemented to balance estimation between accuracy and smoothness.

\subsection{Feature Selection Methodology}
\label{subsec:feature-selection-methodology}
The proposed method consists of two methods in parallel. 

\textbf{Method 1:} is a hybrid method. A filter first ranks candidate features according to their univariate relevance with the steering torque. The ten strongest correlated features are selected based on the absolute Spearman correlation coefficient. The selected features form a new candidate subset for the exhaustive wrapper. Based on ten features, the wrapper evaluates 1023 candidate subsets (models). The chosen training algorithm is a two-state ($K=2$) HMM to balance model complexity with computation time. 

\textbf{Method 2:} performs a sequential forward search (SFS) \citep{liwicki2009} over 18 candidate features (\ref{tab:features}). Starting with an empty "best feature set", new candidate subsets are generated by combining the "best set" with each remaining candidate feature, separately. A candidate feature is kept if it maximizes the performance score of the "best set". The process is terminated either when the next feature does not increase the performance score by more than 1\% or all features are selected. A two-state HMM is also chosen as the training algorithm. 

\subsection{Feature Selection Analysis}
\label{subsec:evaluation-and-selection}
Selecting the optimal feature subset consists of three steps. As a benchmark, a baseline model was trained with all candidate features (\ref{tab:features}) and 2 hidden states.

\textbf{Step 1:} Preliminary results showed that no single best feature subset exists for all validation recordings. Therefore, feature relevance is determined by counting feature occurrences. For each validation recording, the best model is determined by evaluating the Performance Score (PS), 
\begin{equation}
    \label{eq:ps}
    PS = \omega_1 \cdot \lVert\left(100-A_{T,est}\right)\rVert + \omega_2 \cdot \lVert SM_{T,est}\rVert
\end{equation}

where $\omega_1$ and $\omega_2$ are designed to balance estimation accuracy, $A_{T,est}$, with estimation smoothness, $SM_{T,est}$, 
    \begin{equation}
        \label{eq:ps-accuracy}
        A_{T,est} = \left[1-\frac{1}{SD(T_{d})}RMSE_{T,est}\right] \times 100
    \end{equation}
    \begin{equation}
        \label{eq:ps-smoothness}
        SM_{T,est} = \sqrt{\frac{1}{N}\sum_{i=1}^{N}\left(\dot{T}_{i,est}-\bar{\dot{T}}_{est}\right)}
    \end{equation}

with the root-mean-square-error of estimated steering torque, $RMSE_{T,est}$, calculated as, 

    \begin{equation}
        \label{eq:ps-rmset}
        RMSE_{T,est} = \sqrt{\frac{1}{N}\sum_{i=1}^{N}\left(T_{i,est}-T_{i,d}\right)^2}
    \end{equation}

Lower scoring models are considered better. The features in the best set for each recording are counted cumulatively. The higher the count, the more relevant the feature is considered. This process is repeated for both wrapper methods over a range of metric weights to map feature relevance. The goal is to select the metric weights that result in the simplest model with an optimal balance between accuracy and smoothness.

\textbf{Step 2:} Combining the results of both wrapper methods for the selected metric weights, new feature subsets are generated by grouping on feature relevance. To limit the number of subsets, three selection strategies are defined. The new subsets contain...
\begin{itemize}
    \item \underline{Strict Selection}: ...features with $\geq 66\%$ relevance.
    \item \underline{Mild Selection}: ...features with $\geq 33\%$ relevance.
    \item \underline{Liberal Selection}: ...features counted at least once.
\end{itemize}

For each new subset, a two-state HMM is trained. Fixing the weight trade-off in Step 1 allows to directly compare the trained models by their PS. 

\textbf{Step 3:} The lowest scoring model defines the optimal feature subset.

\subsection{State Selection}
\label{subsec:model-complexity}
The model configuration is determined by fixing the feature subset and selecting the number of states $K$ accordingly. Increasing the states allows to capture driver steering torque in greater detail. However, this also increases model dependency on the training data, known as overfitting. As suggested by Faller, evaluating model performance on an (unseen) validation set implicitly ensures generalization capabilities and model complexity. As an extension in this work, the PS (Equation \ref{eq:ps}) provides a means to balance the selection of $K$. New models are trained for a range of states and different weight trade-offs and their PS is evaluated on each validation recording separately. State selection $K$ is determined by the minimum of two methods. The first method selects the states based on the average PS over all recordings. The second method selects the states per recording and averages over all recordings. For both methods, the search is terminated if the addition of another state does not improve the PS by $>$1\%.

\section{Driving Simulator Experiment}
\label{sec:driving-simulator-experiment}
Humans seldom reproduce identical actions when presented with identical scenarios due to their stochastic nature \citep{kolekar2018}. Therefore, a driving experiment was performed at Toyota Motor Europe (TME) to gather sufficient naturalistic driving data for training, validation, and testing of the HMM driver steering torque model.

\subsection{Driving Scenario}
\label{subsec:driving-scenario}
The scenario consists of 200 km driving on a randomly generated three-lane highway (3.5m lane width based on EU regulations) using an experimentally validated vehicle model with enhanced steering dynamics \citep{damian2022}. A real-world height profile was added to reduce average look-ahead distance and increase immersion. To maintain participant concentration, the scenario is split into 24 sections of 5 min ($\sim 8\ km$). Twelve sections were designed to maximize variability and used for model training. The remaining twelve sections provide new driving scenarios used for model validation and testing. Participants were tasked to manually follow the lane center as close to their preferred driving style without secondary tasks. The vehicle speed was kept at 100 km/h with a commercial cruise controller. 

\subsection{Experimental Procedure}
\label{subsec:experimental-procedure}
\begin{figure}[h!]
    \centering
    \includegraphics[width=0.5\linewidth]{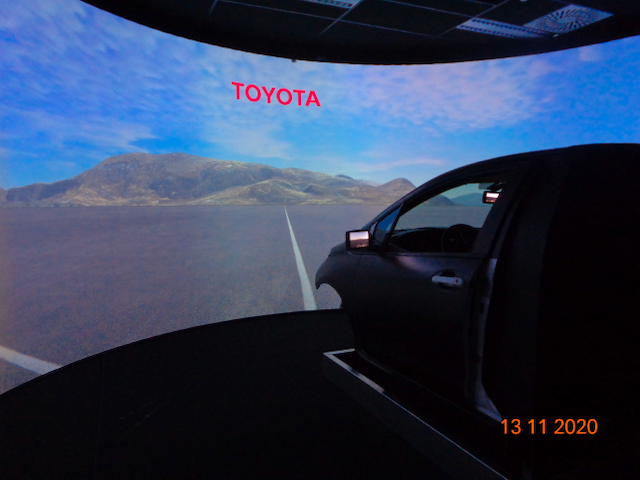}
    \caption{Static driving simulator at Toyota Motor Europe.}
    \label{fig:simulator}
\end{figure}

A total of 7 participants took part in the experiment, all TME staff members involved in DSAS development and testing, averaging 32 years of age (SD = 6.6). Participants, except one, had a driver’s license for an average of 15 years (SD = 6.5). Among the participants were two expert test drivers, two advanced, two intermediate, and one novice driver. The experiment took place in a static driving simulator with a mock-up Toyota production vehicle in front of a 210$^{\circ}$ projection screen, see Fig. \ref{fig:simulator}. The scenarios were rendered with rFpro software. Participants performed two sessions, split over separate days, each in which twelve sections were driven. The order of the trials was randomized for every session and every participant. At the start, participants were able to familiarise themselves with the conditions during a test trial that was not recorded. A total of 14 hours of data were recorded.

%
\section{Results \& Discussion}
\label{sec:results-discussion}

\subsection{Feature Correlation}
Features were ranked based on the average correlation over all training recordings. The ten strongest correlated features ($r_s \geq 0.9$) used in the exhaustive wrapper (sub-method 1) are summarised in Table \ref{tab:feature-correlation}.
Feature relevance of the SFS-wrapper should determine if this is justified. 

\begin{table}[!hbt]
\centering
\captionsetup{width=0.9\columnwidth}
\caption{Feature Correlation}
\label{tab:feature-correlation}
\begin{tabular}{l|l|cc}
 & \textbf{Features} & \textbf{$\lvert r_s \rvert$} & \textbf{$p$} \\ \hline
1 & Steering Wheel Angle & 0.9557 & 0 \\
2 & Deviation Angle @30m & 0.9484 & 0 \\
3 & Road Curvature @10m & 0.9321 & 0 \\
4 & Road Curvature @30m & 0.9251 & 0 \\
5 & Yaw Rate & 0.9249 & 0 \\
6 & Lateral Acceleration & 0.9143 & 0 \\
7 & Roll Angle & 0.9123 & 0 \\
8 & Road Curvature @0m & 0.9099 & 0 \\
9 & Slip Angle & 0.9062 & 0 \\
10 & Lateral Velocity & 0.9058 & 0 \\ \hline\hline
11 & Deviation Angle @10m & 0.8848 & 0 \\ \hline
\end{tabular}
\end{table}

\subsection{Determining Feature Relevance}
\label{subsec:feature-relevance}
\begin{figure}[h!]
    \centering
    \begin{subfigure}[b]{0.99\columnwidth}
        \centering
        \includegraphics[width=0.7\textwidth]{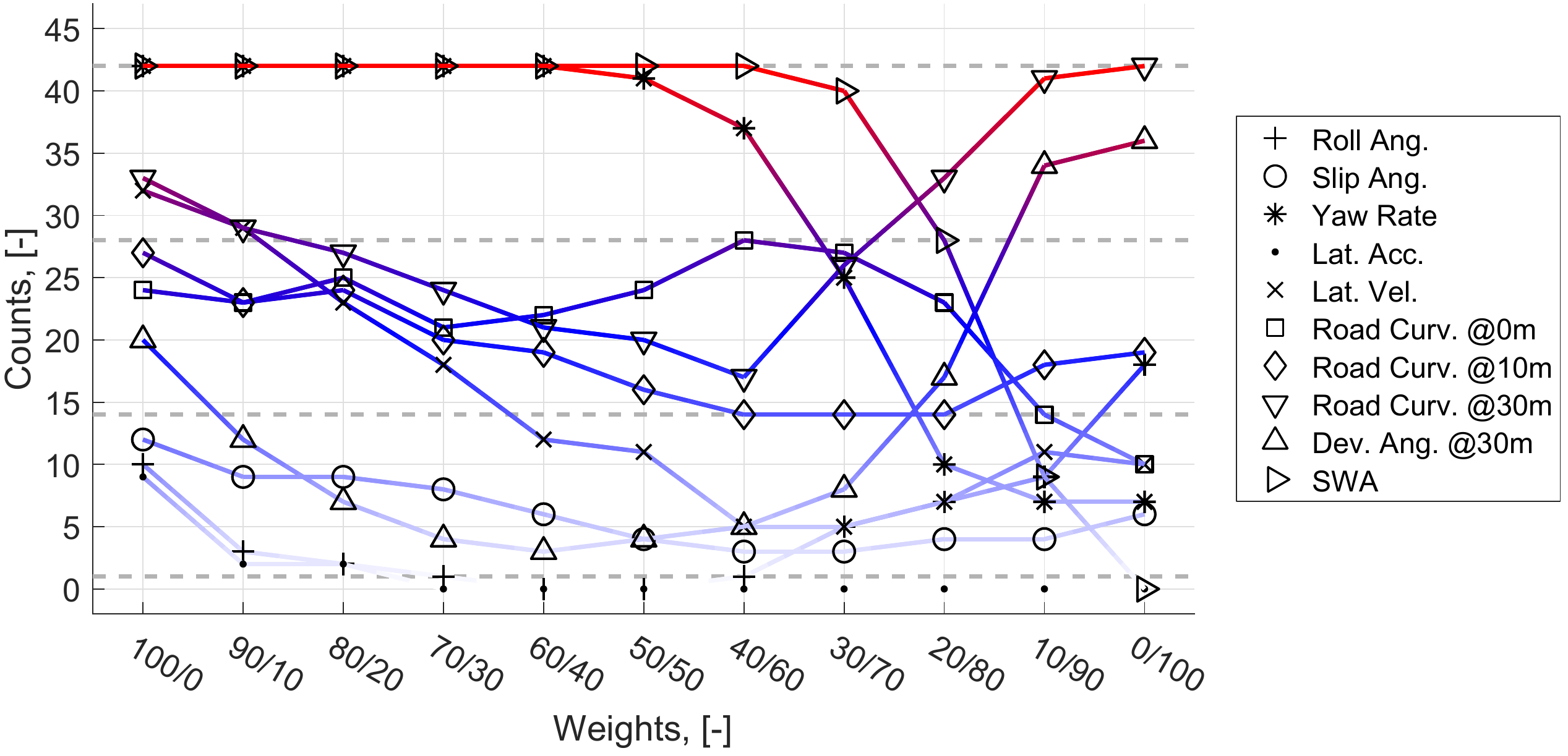}
        \caption{Exhaustive Wrapper}
        \label{fig:feature-relevance-counts1}
    \end{subfigure}
    \begin{subfigure}[b]{0.99\columnwidth}
        \centering
        \includegraphics[width=0.7\textwidth]{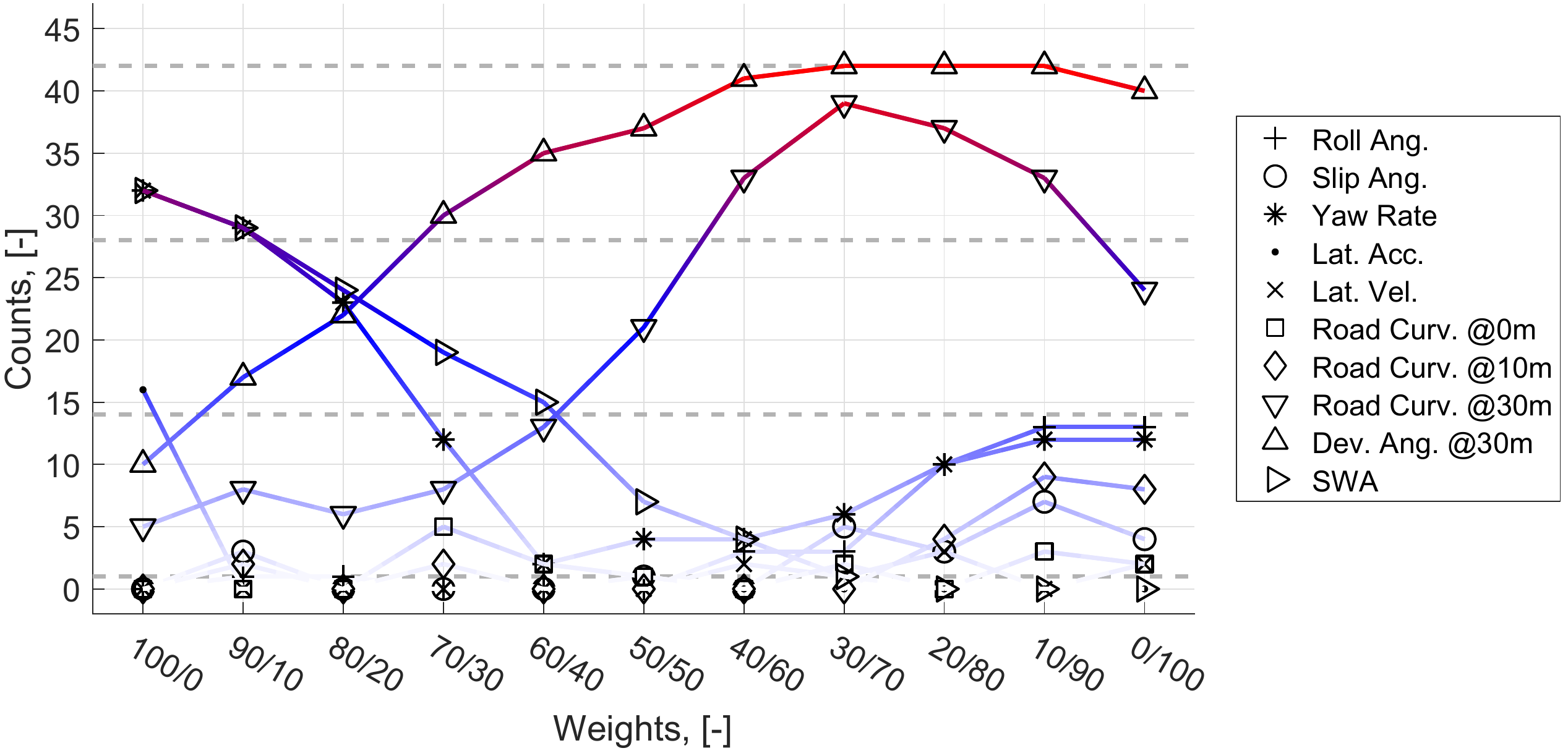}
        \caption{SFS Wrapper (relevant features to Exhaustive Wrapper)}
        \label{fig:feature-relevance-counts2a}
    \end{subfigure}
    \begin{subfigure}[b]{0.99\columnwidth}
        \centering
        \includegraphics[width=0.7\textwidth]{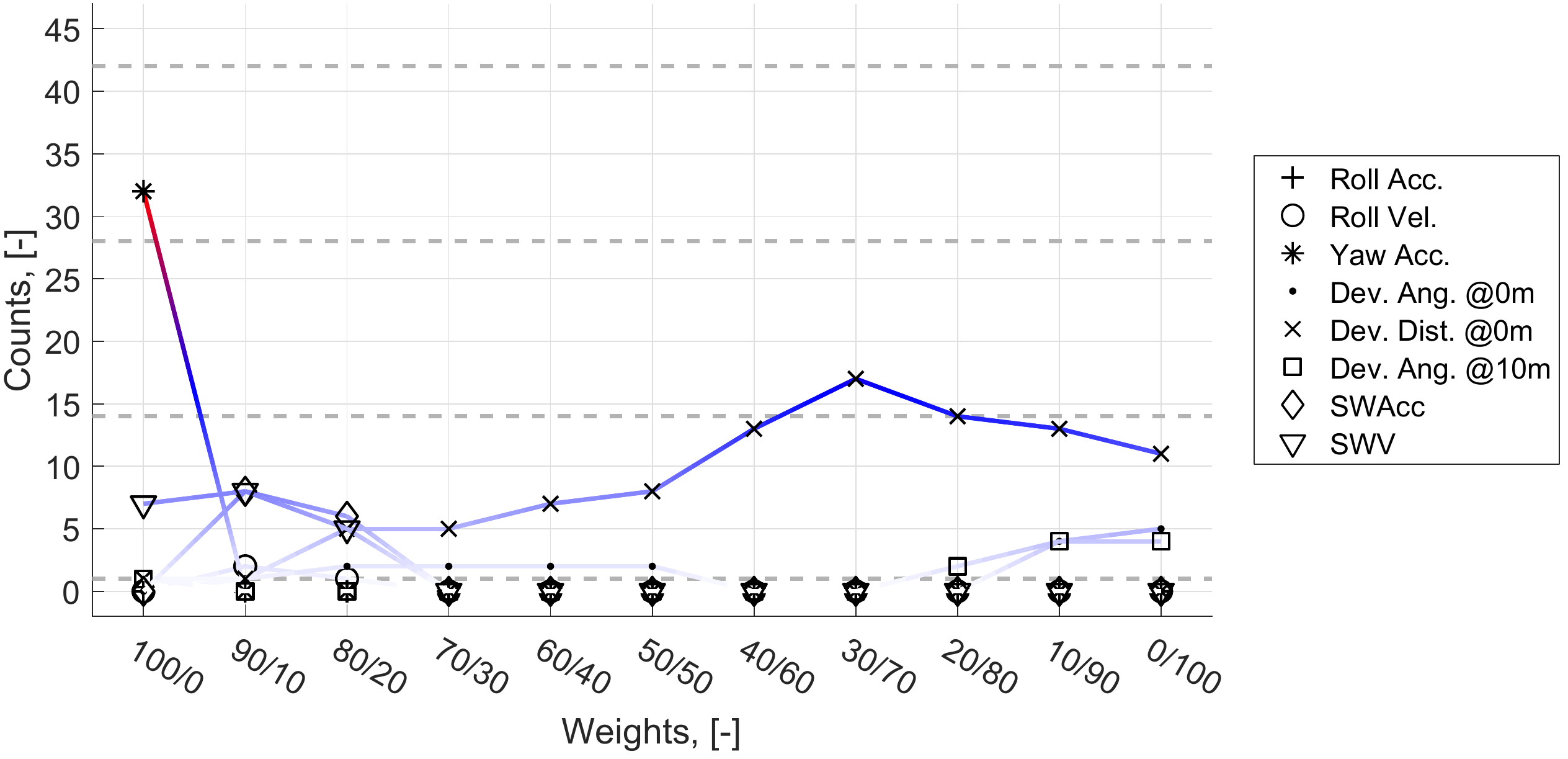}
        \caption{SFS Wrapper (disregarded features)}
        \label{fig:feature-relevance-counts2b}
    \end{subfigure}
    \caption{Mapping influence of metric weights on feature counts. Weights are denoted as accuracy/smoothness.}
    \label{fig:feature-relevance}
\end{figure}

The influence of the weight trade-offs on feature relevance was mapped in ten percent intervals (Figure \ref{fig:feature-relevance}). The maximum count is equal to the number of validation recordings, 42. From Fig. \ref{fig:feature-relevance-counts1} it is observed that for increased estimation smoothness, overall feature relevance decreases up to a balanced trade-off. This means that the average number of features contained in the subsets decreases, indicating simpler models. Furthermore, while vehicle dynamics and driver input are more relevant for accuracy, road preview features become more relevant for smoother estimations. This intuitive outcome is confirmed by the SFS wrapper in Fig. \ref{fig:feature-relevance-counts2a}. However, SFS results show a more conservative count due to the more limited exploration. 
Fig. \ref{fig:feature-relevance-counts2b} confirms the suitability of the filter method to discard features.

\begin{figure}[h!]
    \centering
    \begin{subfigure}[b]{0.45\columnwidth}
        \centering
        \includegraphics[width=\textwidth]{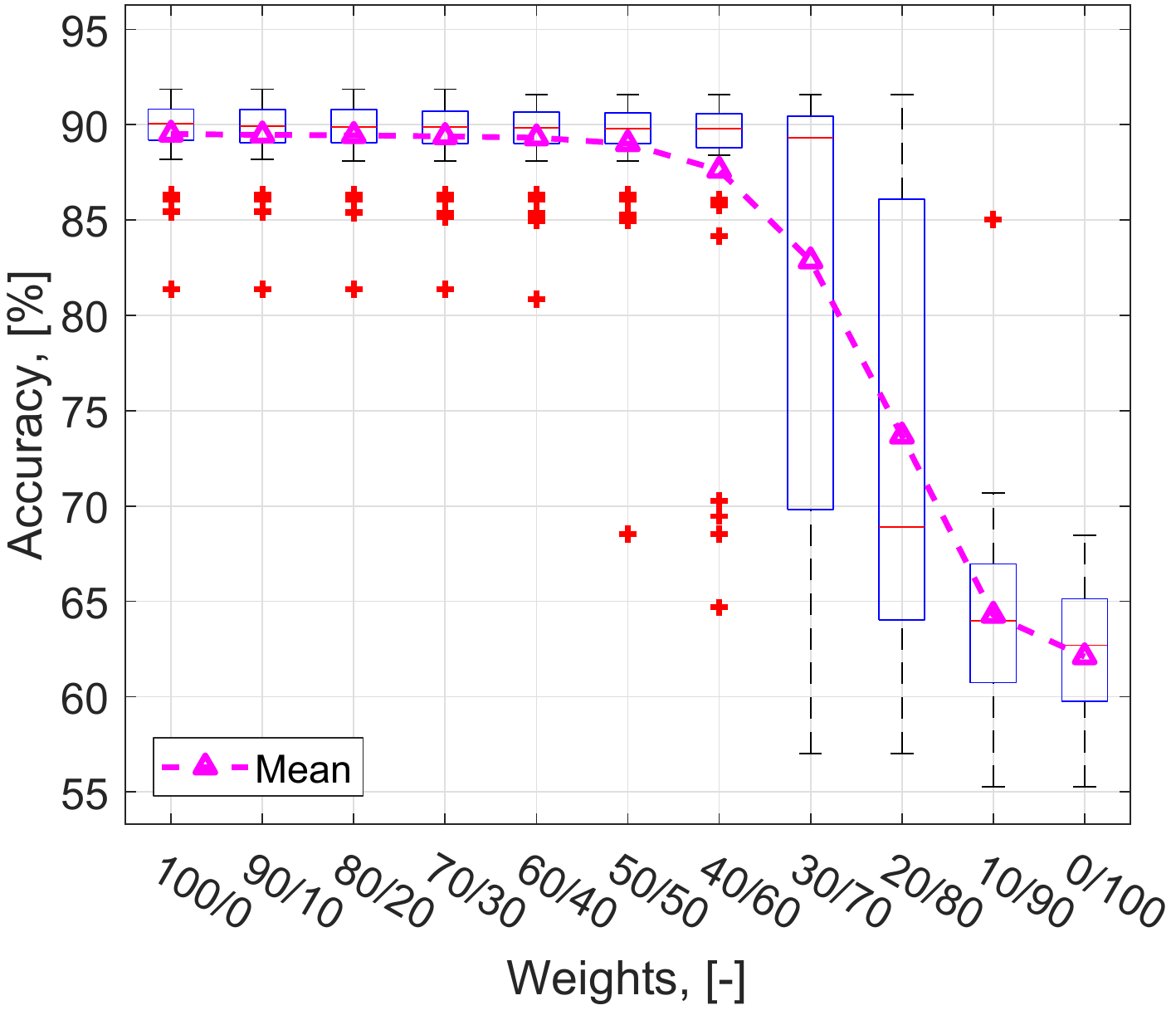}
        \caption{Model accuracy}
        \label{fig:feature-relevance-acc1}
    \end{subfigure}
    \hfill
    \begin{subfigure}[b]{0.45\columnwidth}
        \centering
        \includegraphics[width=\textwidth]{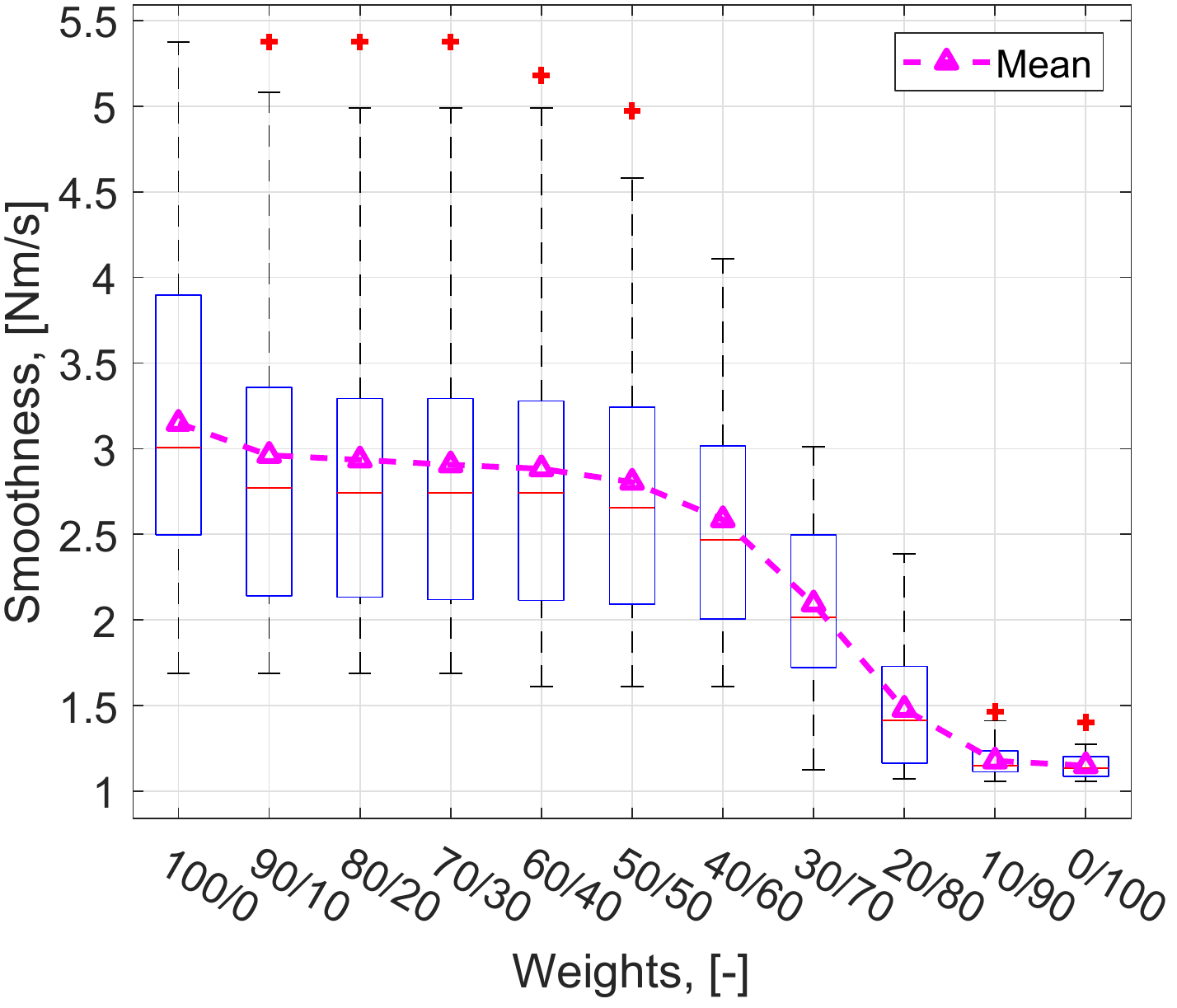}
        \caption{Estimation smoothness}
        \label{fig:feature-relevance-sm1}
    \end{subfigure}
    \caption{Influence of metric weights on model performance for the exhaustive wrapper method.}
    \label{fig:feature-relevance-metrics}
\end{figure}

\begin{figure}[h!]
    \centering
    \includegraphics[width=0.6\columnwidth]{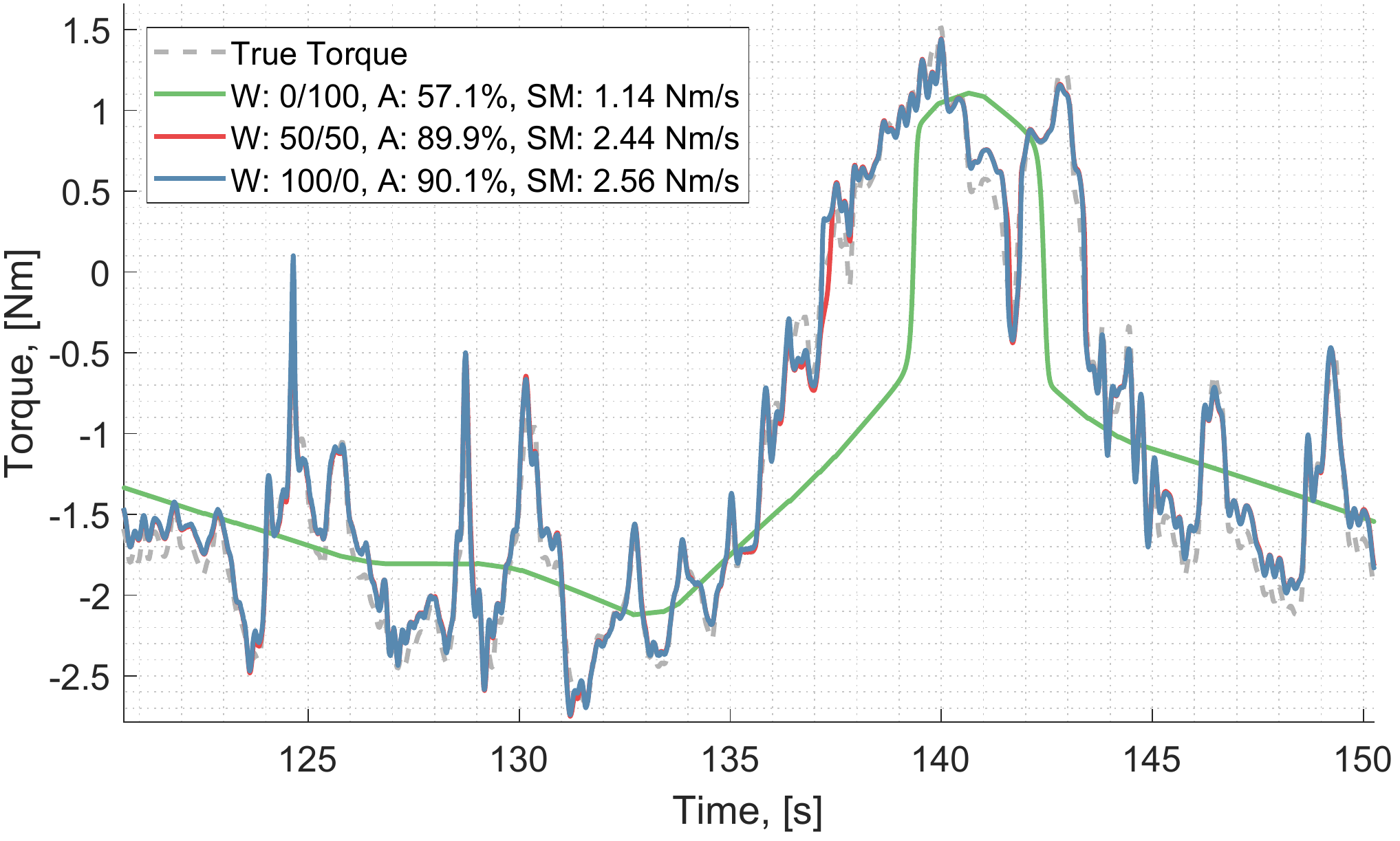}
    \caption{Steering torque estimations of the optimal models for different weight trade-offs.}
    \label{fig:feature-relevance-estimations}
\end{figure}

Moreover, it is observed from Fig. \ref{fig:feature-relevance-metrics} that accuracy and smoothness scores remain approximately constant over the range 100/0 to 40/60, averaging 89\% and 2.94 Nm/s, respectively, while for the same range overall feature relevance decreases. Similar accuracy and smoother estimations are achieved with fewer features and thus simpler models. 
The outliers correspond to the recordings of the "Novice" participant and potentially indicate decreased model performance for novice drivers.
To visualize model performance, Fig. \ref{fig:feature-relevance-estimations} compares different PS weight trade-offs. It shows that the smoothest model lacks accuracy, whereas a pure focus on accuracy increases estimation quality rapidly but at the cost of higher noise. Therefore, a 50/50 weight trade-off is chosen as it objectively balances both metrics, while reducing model complexity.

\subsection{Optimal Feature Subset}
\begin{figure}[h!]
    \centering
    \includegraphics[width=0.45\columnwidth]{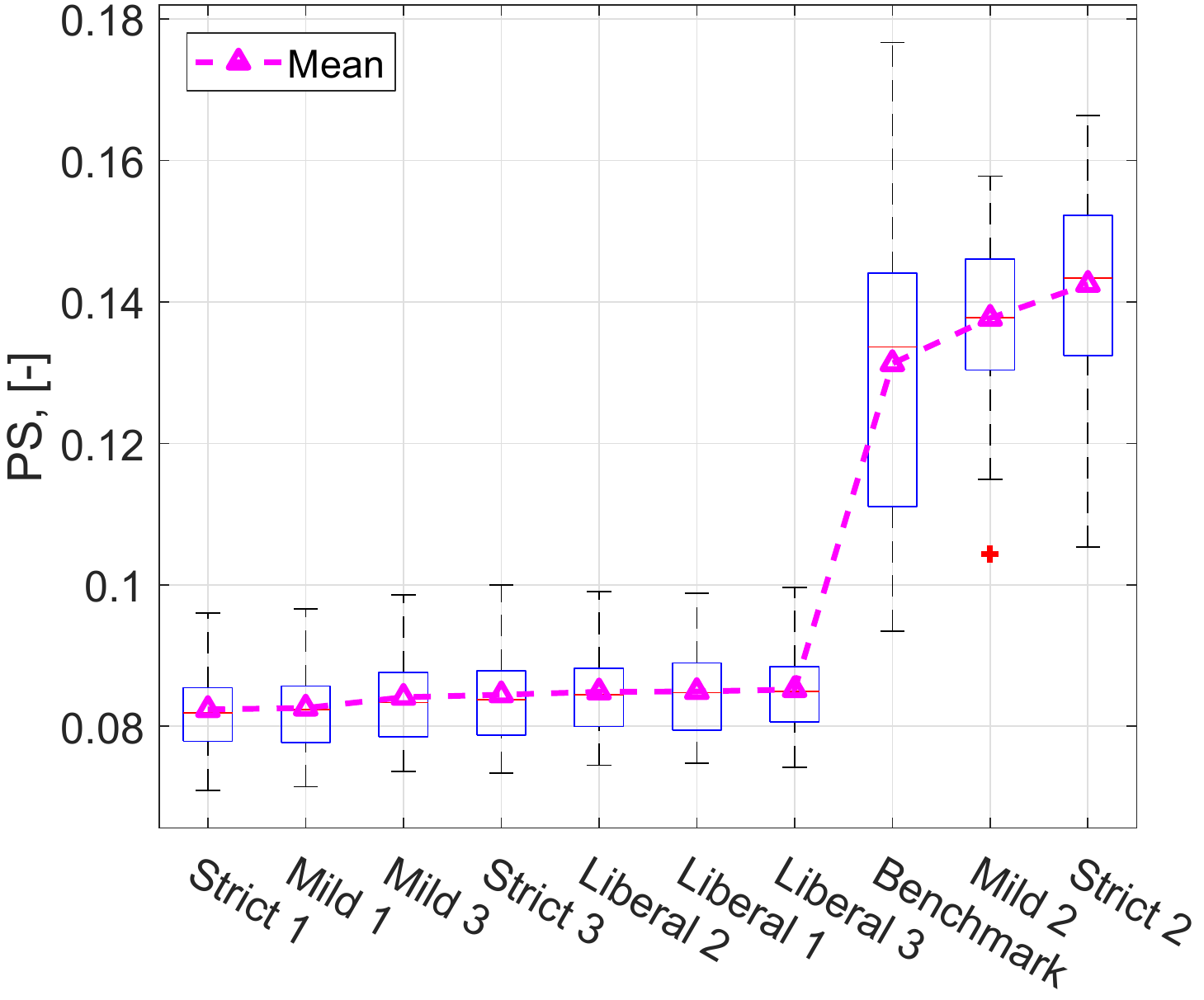}
    \caption{Performance score of subset strategies}
    \label{fig:selection-strategies-metrics}
\end{figure}

Final performance still depends on a specific feature subset. All final subsets combinations are found in Appendix \ref{subsec:selection_strategies}. For each subset, a new 2-state HMMs is trained and validated for a 50/50 weight trade-off in Fig. \ref{fig:selection-strategies-metrics}. The baseline model, defined in \ref{subsec:evaluation-and-selection}, is the most accurate (avg. 91\%) but also the least smooth (avg. 5.82 Nm/s). Except for "Strict 2" and "Mild 2" (producing inaccurate estimations, avg. 62\%), all remaining selection strategies improved upon the baseline model based on their respective performance scores. As each selection was able to achieve similar accuracies (89$\pm$0.08\%), sorting is dictated by the respective smoothness scores. The selection with the least amount of features (steering wheel angle, $\theta_{SWA}$, and yaw rate, $\dot{\psi}$), denoted "Strict 1", can achieve the smoothest estimations (2.86 Nm/s) and therefore the best PS.

\subsection{State Selection}
\begin{figure}[h!]
    \centering
    \begin{subfigure}[b]{0.45\columnwidth}
        \centering
        \includegraphics[width=\textwidth]{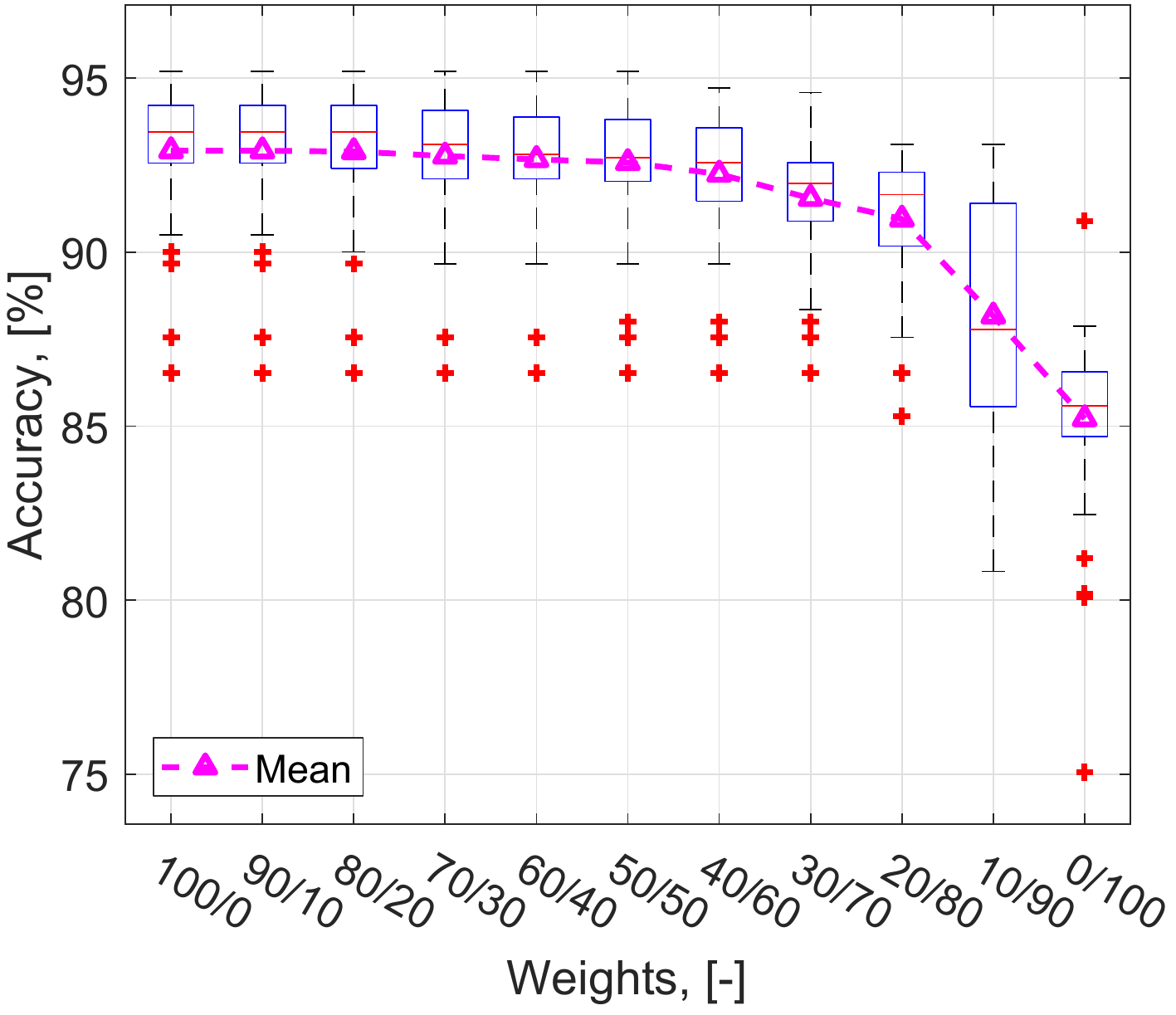}
        \caption{Model accuracy}
        \label{fig:state-selection-acc}
    \end{subfigure}
    \hfill
    \begin{subfigure}[b]{0.45\columnwidth}
        \centering
        \includegraphics[width=\textwidth]{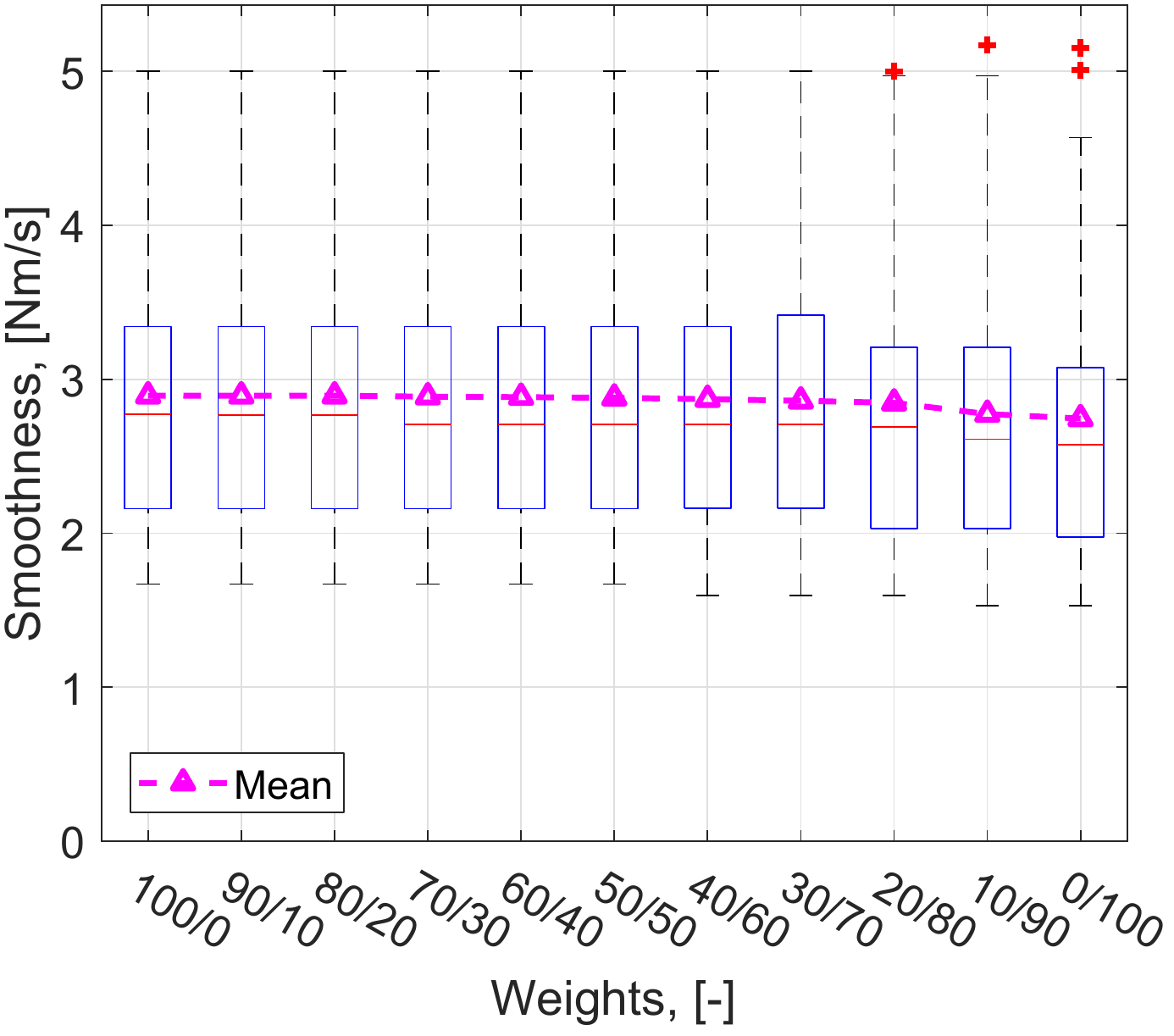}
        \caption{Estimation smoothness}
        \label{fig:state-selection-sm}
    \end{subfigure}
    \hfill
    \begin{subfigure}[b]{0.99\columnwidth}
        \centering
        \includegraphics[width=0.6\textwidth]{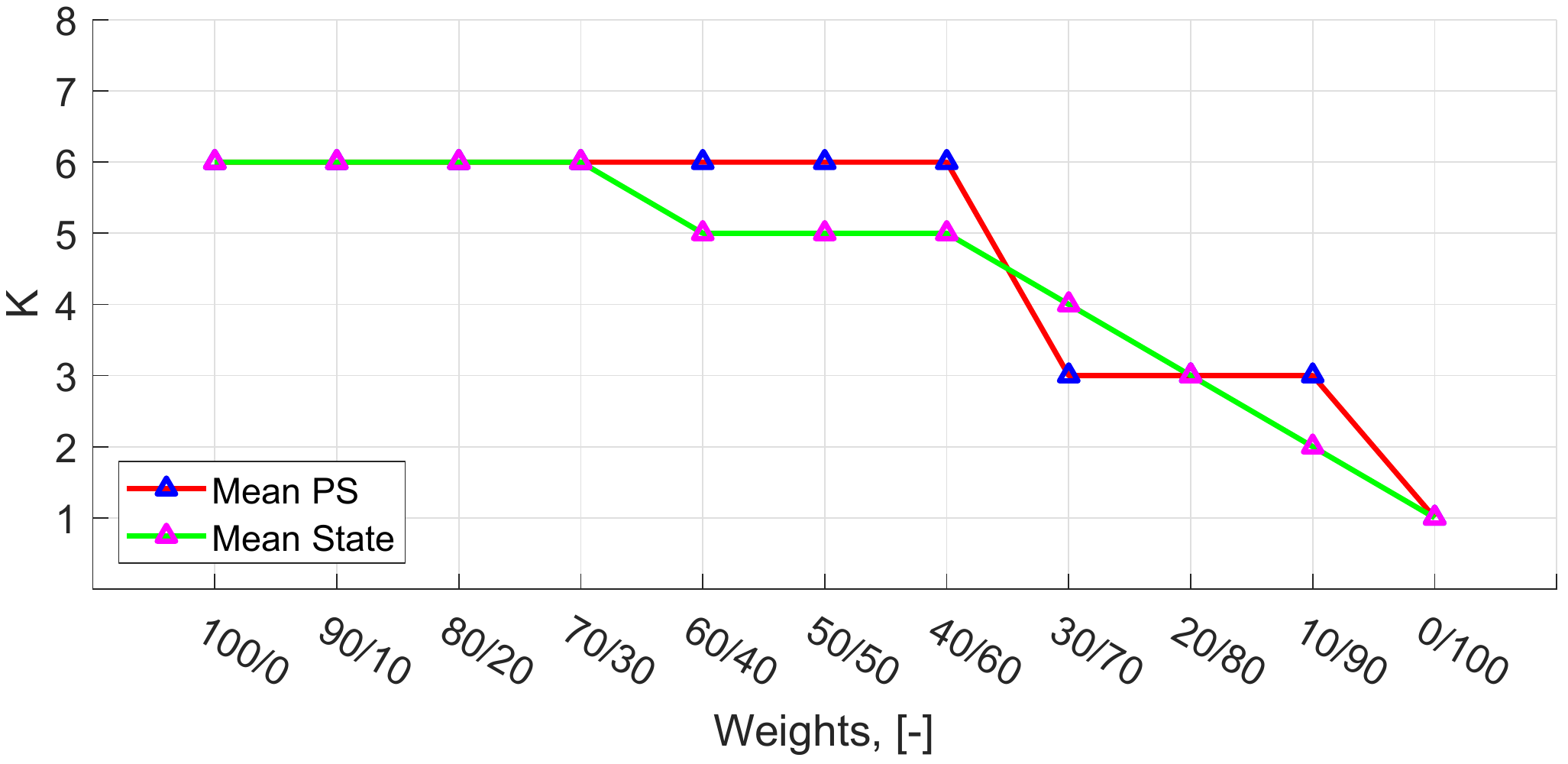}
        \caption{State Selection}
        \label{fig:state-selection-nstates}
    \end{subfigure}
    \caption{Influence of metric weights on state selection for an HMM model with "Strict 1" feature selection.}
    \label{fig:state-selection}
\end{figure}

New models were trained for $K \in \{1,...,20\}$. The influence of metric weights on state selection was mapped in ten percent intervals. Fig. \ref{fig:state-selection} shows that more states are preferred for more accurate models, while fewer states are in favor of smoother estimations. Furthermore, selecting the number of states based on the average $K$ over individual recordings shows a more conservative state selection for most settings and should thus be preferred over the average PS selection. 
As both PS metrics stay virtually constant for weights over 50\%, a balanced weight trade-off is selected. Looking at Fig. \ref{fig:state-selection-nstates}, this results in a selection of five states.

\subsection{Model Testing}
\begin{figure}[h!]
    \centering
    \begin{subfigure}[b]{0.45\columnwidth}
        \centering
        \includegraphics[width=\textwidth]{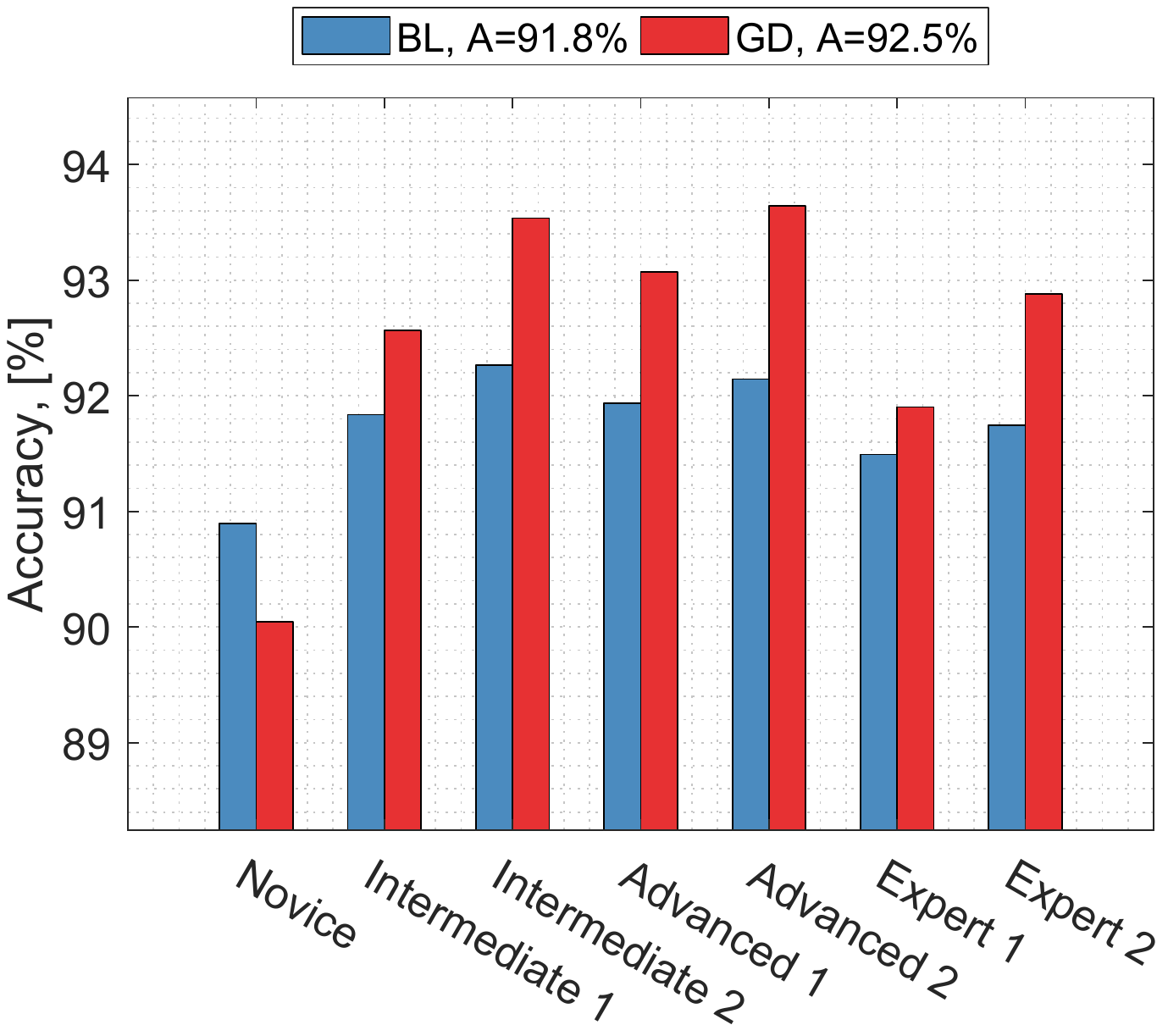}
        \caption{Model accuracy}
        \label{fig:baseline-comparison-acc}
    \end{subfigure}
    \hfill
    \begin{subfigure}[b]{0.45\columnwidth}
        \centering
        \includegraphics[width=\textwidth]{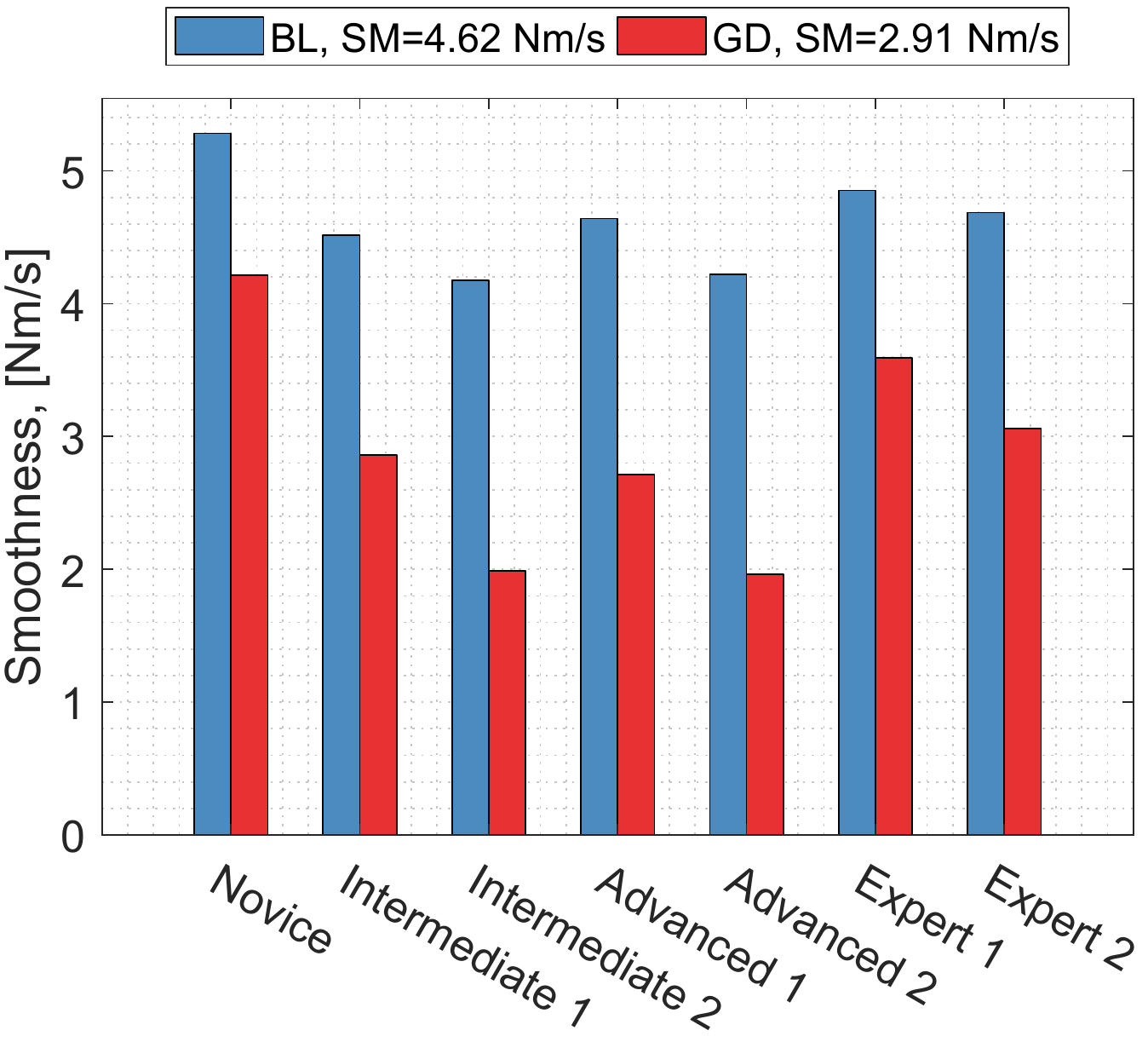}
        \caption{Estimation smoothness}
        \label{fig:baseline-comparison-sm}
    \end{subfigure}
    \hfill
    \begin{subfigure}[b]{0.45\columnwidth}
        \centering
        \includegraphics[width=\textwidth]{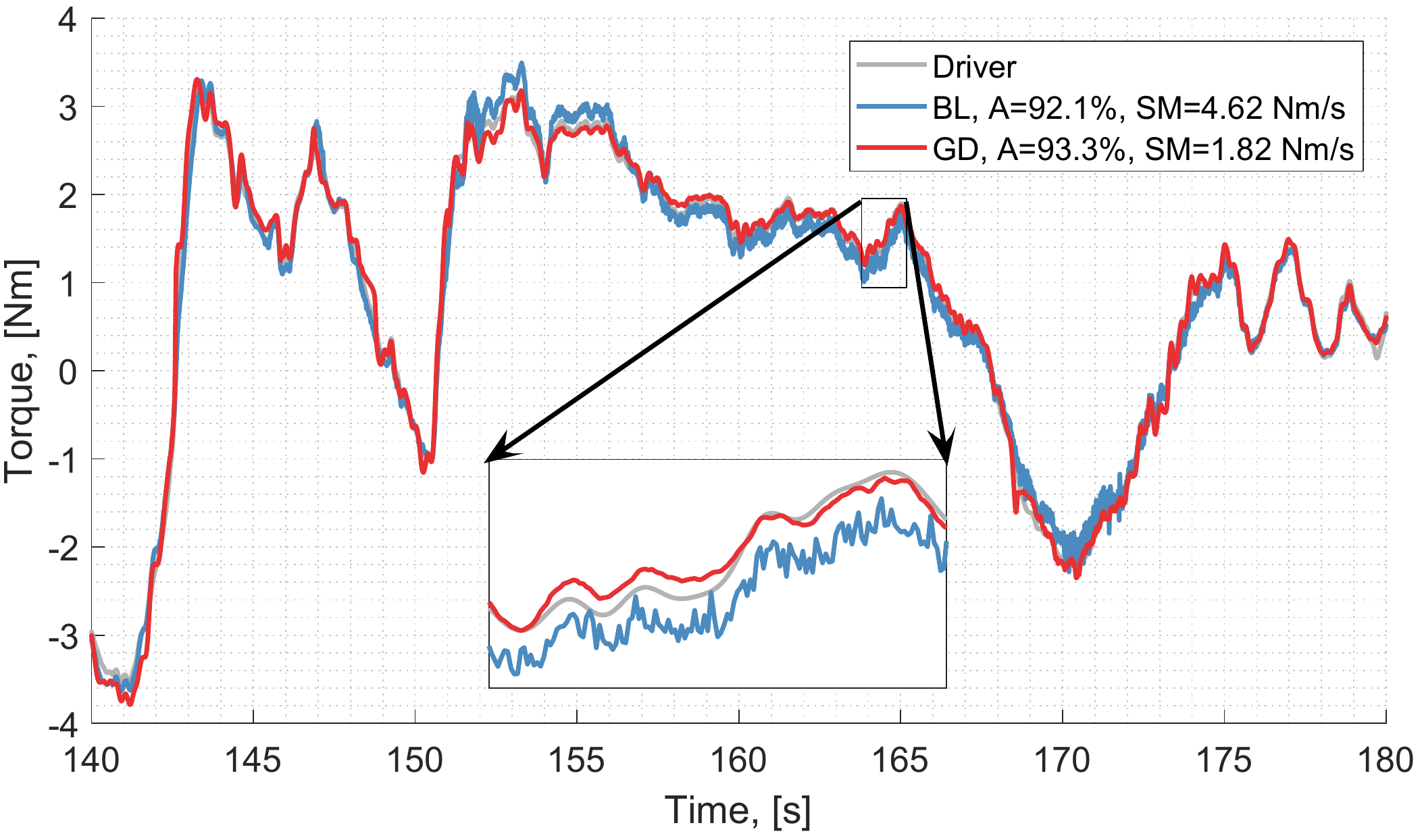}
        \caption{Best performing test-recording}
        \label{fig:baseline-comparison-estimation-a}
    \end{subfigure}
    \hfill
    \begin{subfigure}[b]{0.45\columnwidth}
        \centering
        \includegraphics[width=\textwidth]{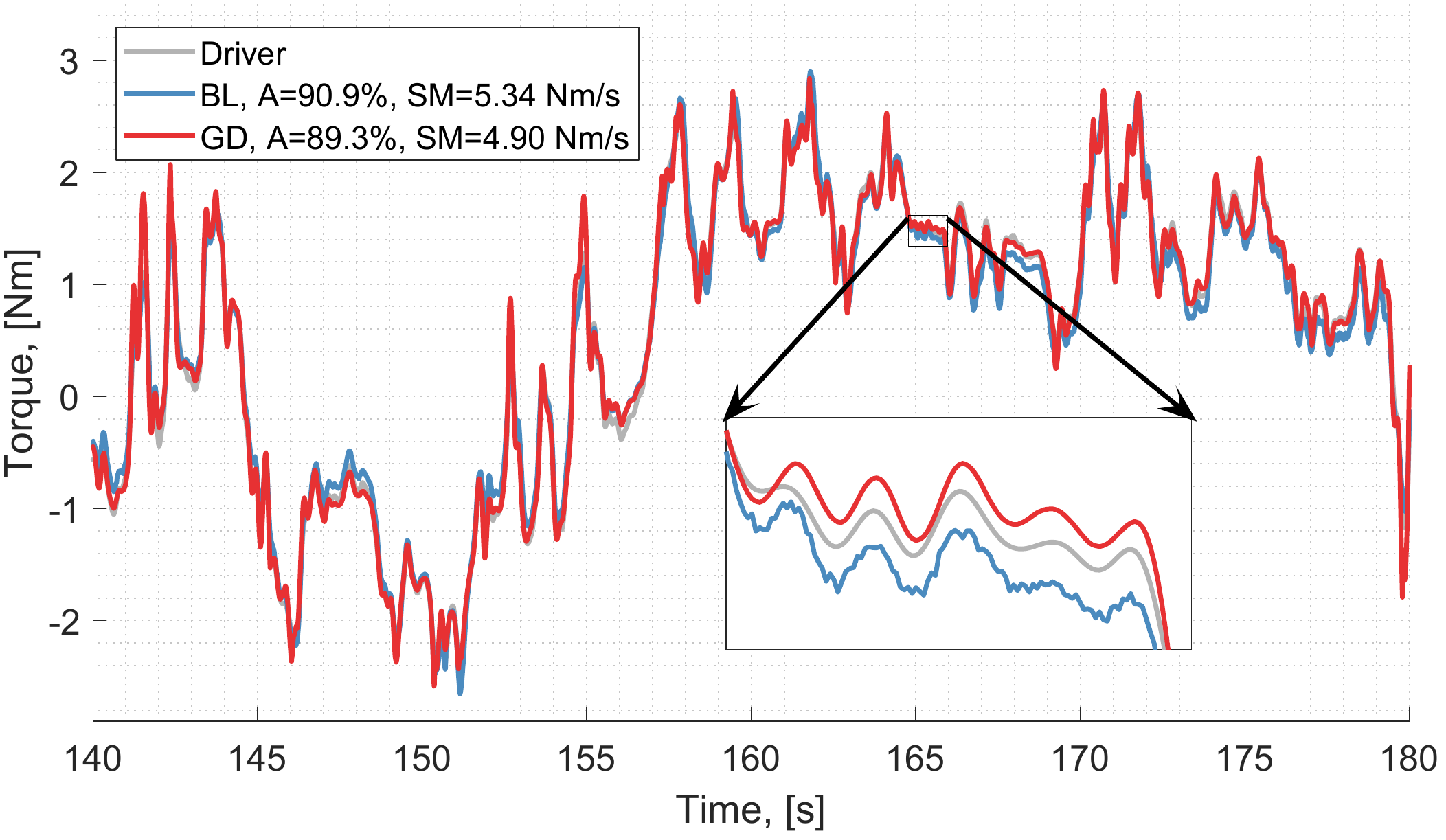}
        \caption{Worst performing test-recording}
        \label{fig:baseline-comparison-estimation-b}
    \end{subfigure}
    \caption{Performance Comparison between Baseline- (BL) and Generic Driver (GD) model.}
    \label{fig:baseline-comparison}
\end{figure}

As a benchmark, a baseline (BL) model was trained with all 18 candidate features and 5 hidden states. Both the Generic Driver (GD) model and BL model were evaluated on the test set. Fig. \ref{fig:baseline-comparison} shows that the GD model can match the BL model's accuracy while simultaneously averaging 37\% smoother estimations. This is best observed in Fig. \ref{fig:baseline-comparison-estimation-a}, where the GD model significantly reduces noise in the estimated steering torque. However, model performance did not improve equally for each participant. Steering behavior of the "Novice" participant was the hardest to estimate both accurately and smoothly, averaging 90\% and 4.22 Nm/s respectively, see Fig. \ref{fig:baseline-comparison-estimation-b}. The reduced model performance can be explained by the more abrupt steering actions of the participant itself. The current participant sample size needs to be increased to generalize conclusions, but performance appears to be dependent on individual steering behavior rather than skill level.

%
%
%
\section{Conclusion}
\label{sec:conclusion}

The proposed HMM-based method addresses the lack of data-driven approaches to model driver steering torque. The results show that our generalized driver model was able to accurately ($\sim 90\%$) estimate driver steering torque while keeping estimations as smooth as possible. Moreover, the model captures complex nonlinear behavior and inter-driver variability from novice to expert drivers. 
In addition, the weighted performance score, allowing a balance between accuracy and smoothness, provides insight to appropriately select the free model parameters while preventing overfitting. 

Finally, the results show an interesting potential to become a vehicle and steering performance predictor in future user-oriented ADAS. In future work, subjective steering feel should be considered in addition to objective metrics. Other directions include online learning approaches for adaptation to individual and time-varying driver behavior, and a wider scope of driving scenarios (lane changes, traffic, off-highway, etc.).

%
%
\bibliography{ifacconf}

%

\appendix
\label{sec:appendix}
%
%
%
\section{List of Selection Strategies based on feature relevance}
\label{subsec:selection_strategies}

\begin{table}[h]
    \centering
     \captionsetup{width=0.9\columnwidth}
     \caption{Selection Strategies}
    \label{tab:selection-strategies}
    \begin{tabular}{r|c|c|c|c|c|c|c|c|c|c}
        \rot{0}{0.2em}{}
        &\rot{45}{0.2em}{$\boldsymbol{\theta_{SWA}}$}
        &\rot{45}{0.2em}{$\boldsymbol{\dot{\psi}}$}
        &\rot{45}{0.2em}{$\boldsymbol{\rho_{0}}$}
        &\rot{45}{0.2em}{$\boldsymbol{\rho_{30}}$}
        &\rot{45}{0.2em}{$\boldsymbol{\rho_{10}}$}
        &\rot{45}{0.2em}{$\boldsymbol{v_y}$}
        &\rot{45}{0.2em}{$\boldsymbol{e_{\psi 30}}$}
        &\rot{45}{0.2em}{$\boldsymbol{\beta}$}
        &\rot{45}{0.2em}{$\boldsymbol{e_{y0}}$}
        &\rot{45}{0.2em}{$\boldsymbol{e_{\psi 0}}$}\\\hline
        \multicolumn{11}{c}{Counts}\\\hline
        Exhaustive&42&41&24&20&16&11&4&4&0&0\\
        SFS&7&4&1&21&0&0&37&1&8&2\\\hline
        \multicolumn{11}{c}{Strategies}\\\hline
        Strict 1&X&X& & & & & & & & \\
        Strict 2& & & & & & &X& & & \\
        Strict 3&X&X& & & & &X& & & \\\hline
        Mild 1&X&X&X&X&X& & & & & \\
        Mild 2& & & &X& & &X& & & \\
        Mild 3&X&X&X&X&X& &X& & & \\\hline
        Liberal 1&X&X&X&X&X&X&X&X& & \\
        Liberal 2&X&X&X&X& & &X&X&X&X\\
        Liberal 3&X&X&X&X&X&X&X&X&X&X\\\hline\hline
    \end{tabular}
\end{table}

%
\section{List of investigated Candidate Features}
\label{subsec:features}

\begin{table}[h!]
\centering
\captionsetup{width=0.9\columnwidth}
\caption{Candidate Features}
\label{tab:features}
\begin{tabular}{lll}
\multicolumn{1}{l}{\textbf{Feature Name}}  & \multicolumn{1}{l}{\textbf{Symbol}}  & \multicolumn{1}{l}{\textbf{Unit}}\\ \hline
\multicolumn{3}{c}{Driver Input}   \\ \hline
\multicolumn{1}{l|}{Steering Wheel Angle}  & \multicolumn{1}{l|}{$\theta_{SWA}$}  & \multicolumn{1}{l}{rad} \\
\multicolumn{1}{l|}{Steering Wheel Velocity}  & \multicolumn{1}{l|}{$\dot{\theta}_{SWA}$}  & \multicolumn{1}{l}{rad/s} \\
\multicolumn{1}{l|}{Steering Wheel Acceleration} & \multicolumn{1}{l|}{$\ddot{\theta}_{SWA}$}  & \multicolumn{1}{l}{rad/s$^2$}  \\ \hline
\multicolumn{3}{c}{Vehicle Dynamics}  \\ \hline
\multicolumn{1}{l|}{Lateral Velocity} & \multicolumn{1}{l|}{$v_y$}  & \multicolumn{1}{l}{m/s} \\
\multicolumn{1}{l|}{Lateral Acceleration}   & \multicolumn{1}{l|}{$a_y$}  & \multicolumn{1}{l}{m/s$^2$}   \\
\multicolumn{1}{l|}{Slip Angle} & \multicolumn{1}{l|}{$\beta$}   & \multicolumn{1}{l}{rad}  \\
\multicolumn{1}{l|}{Yaw Rate}   & \multicolumn{1}{l|}{$\dot{\psi}$}  & \multicolumn{1}{l}{rad/s} \\
\multicolumn{1}{l|}{Yaw Acceleration} & \multicolumn{1}{l|}{$\ddot{\psi}$} & \multicolumn{1}{l}{rad/s$^2$}   \\
\multicolumn{1}{l|}{Roll Angle} & \multicolumn{1}{l|}{$\phi$} & \multicolumn{1}{l}{rad} \\
\multicolumn{1}{l|}{Roll Velocity} & \multicolumn{1}{l|}{$\dot{\phi}$} & \multicolumn{1}{l}{rad/s}  \\
\multicolumn{1}{l|}{Roll Acceleration}   & \multicolumn{1}{l|}{$\ddot{\phi}$} & \multicolumn{1}{l}{rad/s$^2$}   \\ \hline
\multicolumn{3}{c}{Road Preview}   \\ \hline
\multicolumn{1}{l|}{Deviation Distance @0m} & \multicolumn{1}{l|}{$e_{y0}$}  & \multicolumn{1}{l}{m}   \\
\multicolumn{1}{l|}{Deviation Angle @0m} & \multicolumn{1}{l|}{$e_{\psi 0}$} & \multicolumn{1}{l}{rad} \\
\multicolumn{1}{l|}{Road Curvature @0m}  & \multicolumn{1}{l|}{$\rho_{0}$}   & \multicolumn{1}{l}{1/m} \\
\multicolumn{1}{l|}{Deviation Angle @10m}   & \multicolumn{1}{l|}{$e_{\psi 10}$}   & \multicolumn{1}{l}{rad} \\
\multicolumn{1}{l|}{Road Curvature @10m} & \multicolumn{1}{l|}{$\rho_{10}$}  & \multicolumn{1}{l}{1/m} \\
\multicolumn{1}{l|}{Deviation Angle @30m}   & \multicolumn{1}{l|}{$e_{\psi 30}$}   & \multicolumn{1}{l}{rad} \\
\multicolumn{1}{l|}{Road Curvature @30m} & \multicolumn{1}{l|}{$\rho_{30}$}  & \multicolumn{1}{l}{1/m} \\ \hline
\end{tabular}
\end{table}

%
%
\end{document}